\documentclass[prl,%
twocolumn,%
tightenlines,showpacs,%
nofootinbib,%
superscriptaddress,%
amsfonts,amsmath]{revtex4}

\usepackage{graphicx}
\usepackage{bm}

\begin{document}

\title{Improving relativistic MOND with Galileon k-mouflage}

\author{Eugeny~Babichev}
\affiliation{Laboratoire de Physique Th\'eorique d'Orsay,
B\^atiment 210, Universit\'e Paris-Sud 11,
F-91405 Orsay Cedex, France}
\affiliation{AstroParticule \& Cosmologie,
UMR 7164-CNRS, Universit\'e Denis Diderot-Paris 7,
CEA, Observatoire de Paris,
10 rue Alice Domon et L\'eonie
Duquet, F-75205 Paris Cedex 13, France}

\author{C\'edric~Deffayet}
\affiliation{AstroParticule \& Cosmologie,
UMR 7164-CNRS, Universit\'e Denis Diderot-Paris 7,
CEA, Observatoire de Paris,
10 rue Alice Domon et L\'eonie
Duquet, F-75205 Paris Cedex 13, France}
\affiliation{${\mathcal{G}}{\mathbb{R}}
\varepsilon{\mathbb{C}}{\mathcal{O}}$, Institut d'Astrophysique
de Paris, UMR 7095-CNRS, Universit\'e Pierre et Marie
Curie-Paris 6, 98bis boulevard Arago, F-75014 Paris, France}

\author{Gilles~\surname{Esposito-Far\`ese}}
\affiliation{${\mathcal{G}}{\mathbb{R}}
\varepsilon{\mathbb{C}}{\mathcal{O}}$, Institut d'Astrophysique
de Paris, UMR 7095-CNRS, Universit\'e Pierre et Marie
Curie-Paris 6, 98bis boulevard Arago, F-75014 Paris, France}

\begin{abstract}
We propose a simple field theory reproducing the MOND
phenomenology at galaxy scale, while predicting negligible
deviations from general relativity at small scales thanks to an
extended Vainshtein (``k-mouflage'') mechanism induced by a
covariant Galileon-type Lagrangian. The model passes solar-system
tests at the post-Newtonian order, including those of local
Lorentz invariance, and its anomalous forces in binary-pulsar
systems are orders of magnitude smaller than the tightest
experimental constraints. The large-distance behavior is obtained
as in Bekenstein's tensor-vector-scalar (TeVeS) model, but with
several simplifications. In particular, no fine-tuned function is
needed to interpolate between the MOND and Newtonian regimes, and
no dynamics needs to be defined for the vector field because
preferred-frame effects are negligible at small distances. The
field equations depend on second (and lower) derivatives, and
avoid thus the generic instabilities related to higher
derivatives. Their perturbative solution around a Schwarzschild
background is remarkably simple to derive. We also underline why
the proposed model is particularly efficient within the class of
covariant Galileons.
\end{abstract}

\date{\today}


\maketitle


Although the hypothesis of dark matter is consistent with a wide
range of observations, it might be an artifact of our Newtonian
interpretation of experimental data if the gravitational $1/r^2$
law happens not to be valid at large distances. In 1983, Milgrom
proposed a simple phenomenological rule, called ``modified
Newtonian dynamics'' (MOND) \cite{Milgrom:1983ca}, depending on
an acceleration scale $a_0$: In the vicinity of a (baryonic) mass
$M$, any test particle undergoes the standard Newtonian
acceleration $a_N = GM/rc^2$ if $a_N > a_0$, but a slower
decreasing one $a = \sqrt{a_0 a_N} = \sqrt{GMa_0}/r$ if $a <
a_0$. This law happens to fit remarkably well galaxy rotation
curves for a universal constant~\cite{Sanders:2002pf}
\begin{equation}
a_0 \approx 1.2 \times 10^{-10}\, \text{m.s}^{-2}.
\label{a0}
\end{equation}
Moreover, it automatically recovers the Tully-Fisher
law~\cite{Tully:1977fu} and explains the correlation of dark
matter profiles with the baryonic ones \cite{McGaugh:2005er}.

However, reproducing this phenomenological law in a consistent
relativistic field theory happens to be quite difficult, as
illustrated by almost three decades of abundant literature. In
the present paper, we shall refine on one of the best models
proposed so far, in our opinion, namely the tensor-vector-scalar
(TeVeS) theory constructed by Bekenstein and
Sanders~\cite{Bekenstein:1992pj,Bekenstein:1993fs,Sanders:1996wk,
Bekenstein:2004ne,Bekenstein:2004ca,Sanders:2005vd}. We will show
that a generalized Galileon action allows us to suppress
deviations from general relativity at small distances, thanks to
an extension of the Vainshtein mechanism occurring in massive
gravity \cite{Vainshtein:1972sx,Babichev:2010jd}. In the model we
propose, the magnitude itself of the anomalous force tends
towards zero at small distances, and not only its ratio to the
gravitational one (as in the standard Vainshtein mechanism).

We consider a scalar-tensor theory of gravity defined by the
action
\begin{eqnarray}
S&=& \frac{c^3}{4 \pi G}\int d^4 x \sqrt{-g}
\biggl( \frac{R}{4}
+\mathcal{L}_\text{standard}
+\mathcal{L}_\text{MOND}\nonumber\\
&&+\mathcal{L}_\text{Galileon} \biggr)
+ S_\text{matter}\left[\psi_\text{matter};
\tilde{g}_{\mu \nu}\right],
\label{action}
\end{eqnarray}
where we use the sign convention of \cite{MTW} and notably the
mostly plus signature. The physical metric $\tilde{g}_{\mu \nu}$
will be defined in Eq.~(\ref{gtilde}) below. The scalar field
$\varphi$ is chosen dimensionless, and its kinetic term is the
sum of the following three contributions, where $s \equiv g^{\mu
\nu} \varphi_{,\mu}\varphi_{,\nu}$~:
\begin{eqnarray}
\mathcal{L}_\text{standard} &=& -\frac{\epsilon}{2}\,s
= - \frac{\epsilon}{2} (\partial_\lambda\varphi)^2,
\label{standard}\\
\mathcal{L}_\text{MOND} &=& - \frac{c^2}{3a_0} s \sqrt{|s|}\,,
\label{MOND}\\
\mathcal{L}_\text{Galileon} &=& - \frac{k}{3}
\varepsilon^{\alpha \beta \gamma \delta}
\varepsilon^{\mu \nu \rho \sigma}
\varphi_{,\alpha} \varphi_{,\mu} \varphi_{;\beta \nu}
R_{\gamma \delta \rho \sigma}.
\label{Galileon}
\end{eqnarray}
Here $\varepsilon^{\alpha \beta \gamma \delta}$ denotes the
Levi-Civita tensor, related to the fully antisymmetrical symbol
$\left[\alpha \beta \gamma \delta\right]$ (whose values are 0 or
$\pm 1$) by $\varepsilon^{\alpha \beta \gamma \delta}\equiv
(-g)^{-1/2}\left[\alpha \beta \gamma \delta\right]$. A small mass
term $-\frac{1}{2}m^2\varphi^2$, with $1/m$ greater than the
largest cluster sizes, might also be added to the above kinetic
term.

Denoting as $r_\text{MOND} \equiv \sqrt{GM/a_0}$ the scale where
MOND effects start to manifest around a galaxy of baryonic mass
$M$, and as $r_\text{V} \equiv (8kGM a_0)^{1/4}/c <
r_\text{MOND}$ the Vainshtein radius below which scalar effects
are suppressed, we will see below that
$\mathcal{L}_\text{standard}$ dominates at very large distances
$r > r_\text{MOND}/\epsilon$, $\mathcal{L}_\text{MOND}$ at
intermediate ones $r_\text{V} < r < r_\text{MOND}/\epsilon$, and
$\mathcal{L}_\text{Galileon}$ at small scales $r < r_\text{V}$.
As stated above, numerical fits of galaxy rotation curves
\cite{Sanders:2002pf} give the value (\ref{a0}) for $a_0$, while
their flatness up to about $10\, r_\text{MOND}$
\cite{Gentile:2006hv} imposes that the positive dimensionless
constant $\epsilon$ is smaller than $0.1$. The constant $k$
entering $\mathcal{L}_\text{Galileon}$ has dimension of a
[length]${}^4$, and we will show below that a numerical value
\begin{equation}
k \approx \left(4\times 10^{-6}\, \frac{c^2}{a_0}\right)^4
\approx \left(100\, \text{kpc}\right)^4
\label{k}
\end{equation}
allows the model to pass solar-system tests while predicting MOND
effects even for the lightest known dwarf galaxies.

Several ingredients of action (\ref{action}) are borrowed from
the MOND literature. In particular, the scalar kinetic term
$\mathcal{L}_\text{MOND}$ is well known to generate an extra
acceleration $\sqrt{G M a_0}/r$ on any test mass at distance $r$
{}from a source of baryonic mass $M$ \cite{Bekenstein:1984tv}.
Note that we did not write it simply in terms of $|s|^{3/2}$, but
that an absolute value is involved only within the square root,
to ensure that the scalar field carries positive energy whatever
the sign of $s$~\cite{Bruneton:2007si}. The standard kinetic term
$\mathcal{L}_\text{standard}$ has a negligible influence because
it is multiplied by the small positive constant $\epsilon$, but
it ensures that the dynamics of the scalar field is well defined
when $s$ passes through a vanishing
value~\cite{Bruneton:2007si,Babichev:2007dw}.

As in Refs.~\cite{Bekenstein:2004ne,Bekenstein:2004ca,
Sanders:1996wk,Sanders:2005vd}, the matter action
$S_\text{matter}$ assumes that all matter fields $\psi$ are
minimally coupled to a physical metric
\begin{equation}
\tilde{g}_{\mu \nu}\equiv
e^{-2\varphi}g_{\mu\nu}-2\,\text{sinh}(2\varphi)U_\mu U_\nu\,,
\label{gtilde}
\end{equation}
where $U_\mu$ is a timelike unit vector field, i.e.,
$g^{\mu\nu}U_\mu U_\nu = -1$. Light deflection by galaxies and
clusters would indeed be inconsistent with experiment if matter
was merely coupled to the scalar field via a conformal metric
$\tilde{g}_{\mu \nu} = e^{2\varphi}g_{\mu\nu}$
\cite{Bekenstein:1992pj,Bekenstein:1993fs,Bruneton:2007si}. In a
locally inertial frame where $g_{\mu\nu} =
\text{diag}(-1,1,1,1)$, and choosing the observer's velocity such
that $U_\mu = (1,0,0,0)$ lies along his proper time direction,
Eq.~(\ref{gtilde}) merely means that $\tilde g_{00} =
e^{2\varphi}g_{00}$ but $\tilde g_{ij} = e^{-2\varphi}g_{ij}$
(mimicking thus the behavior of pure general relativity in
presence of dark matter). Previous field theories attempting at
reproducing the MOND dynamics predicted large preferred-frame
effects in the solar system, inconsistent with experiment, if
$U_\mu$ was assumed to be a constant vector field
\cite{Sanders:1996wk,Bruneton:2007si}. This is why it was assumed
to be dynamical in the TeVeS model
\cite{Bekenstein:2004ne,Bekenstein:2004ca}, with the idea that it
could align with the matter's local proper time direction.
However, making the vector field dynamical by adding a kinetic
term $-F_{\mu\nu}^2$, or any function of it, causes this vector
to be unstable
\cite{Clayton:2001vy,Bruneton:2007si,EspositoFarese:2009aj}. In
the present paper, we will see that preferred-frame effects
remain negligible in the solar system even if $U_\mu$ is assumed
to be constant, thanks to the Vainshtein mechanism at small
distances. This mechanism also allows us to choose other forms
than (\ref{gtilde}), in contrast to
Refs.~\cite{Bekenstein:2004ne,
Bekenstein:2004ca,Sanders:1996wk,Sanders:2005vd}. We could also
choose the physical metric as $\tilde{g}_{\mu \nu} \approx
e^{2\varphi} g_{\mu\nu}
+B(\varphi,s)\varphi_{,\mu}\varphi_{,\nu}$, but $B$ would need to
be a fine-tuned function of both $\varphi$ and its standard
kinetic term, $s = (\varphi_{,\lambda})^2$, in order to be
consistent with the observed light deflection by galaxies and
clusters. Moreover, the conditions for consistency of the theory
within matter would be quite involved~\cite{Bruneton:2007si}.
The choice (\ref{gtilde}), borrowed from
\cite{Bekenstein:2004ne,Bekenstein:2004ca,
Sanders:1996wk,Sanders:2005vd}, is thus the most natural one in
the present framework.

The reason why a mere action $\mathcal{L}_\text{standard} +
\mathcal{L}_\text{MOND}$, Eqs.~(\ref{standard}) and (\ref{MOND})
above, does not suffice to define a consistent relativistic field
theory of MOND is that it would also predict an extra force
$\sqrt{GM_\odot a_0}/r$ within the solar system (where $M_\odot$
denotes the mass of the Sun), in addition to the Newtonian one
$GM_\odot /r^2$ and its post-Newtonian corrections. [We often use
the word ``force'' instead of ``acceleration'' in the present
paper, i.e., do not write the mass of the test particle to
simplify.] This would be ruled out by tests of Kepler's third law
and those of post-Newtonian dynamics. The literature considered
thus ``Relativistic AQUAdraric Lagrangians'' (RAQUAL), also known
as ``k-essence'' theories in the cosmological framework, i.e., a
scalar kinetic term $-\frac{1}{2}\,f(s)$ depending on a
\textit{function} of $s = (\varphi_{,\lambda})^2$. In order to
reproduce the MOND dynamics, this kinetic term was assumed to
take the form (\ref{MOND}) for small accelerations (i.e., small
values of $s$), while the scalar field behavior within the solar
system depended on the shape of $f(s)$ for large $s$. The clever
choice of the literature was to impose that $f'(s)$ tends towards
a constant value for large $s$, say $f'_\infty$, in order to
recover a Brans-Dicke behavior $\varphi = -GM_\odot/f'_\infty
rc^2$ at small distances, so that the physical metric
(\ref{gtilde}) reproduce the standard Schwarzschild solution up
to a rescaling of the gravitational constant $G_\text{eff} = G
(1+1/f'_\infty)$. The parametrized post-Newtonian (PPN)
parameters $\beta$ and $\gamma$ \cite{Will:2005va} take then
strictly their general relativistic values $\beta = \gamma = 1$,
and classical solar-system tests are passed. However,
binary-pulsar tests are directly sensitive to the matter-scalar
coupling strength $1/f'_\infty$, independently of the fact that
the physical metric (\ref{gtilde}) contains a ``disformal''
contribution proportional to $U_\mu U_\nu$, and they impose
$f'_\infty > 10^4$ \cite{EspositoFarese:2004tu}. As discussed in
\cite{Bruneton:2007si}, such a large value is difficult to
reconcile with the expression (\ref{MOND}) needed for small $s$.
One needs a fine-tuned interpolating function $f(s)$, of a shape
similar to Fig.~3 of \cite{Bruneton:2007si}, in order to predict
MOND effects while passing binary-pulsar tests. Another idea
would thus be to choose a function $f(s)$ such that the scalar
force at small distances is negligible with respect to the
Newtonian one $GM_\odot/r^2$, instead of keeping the same radial
dependence $\varphi'(r) c^2 = GM_\odot/f'_\infty r^2$. As
underlined in \cite{Babichev:2009ee}, nonlinear kinetic terms
(i.e., those of k-essence/RAQUAL models) can reduce scalar
effects at small distances, acting as a camouflage for the scalar
(hence the name ``k-mouflage''). However, RAQUAL models must
satisfy two conditions to have a Hamiltonian bounded by below and
a well-posed Cauchy problem~\cite{Bruneton:2007si}: $f'(s) > 0$
and $2 s f''(s) + f'(s) > 0$. These conditions suffice to prove
that $\varphi''(r) < 0$, i.e., that $\varphi'(r)$ is a decreasing
function of $r$, and the best we can obtain is thus an almost
constant force $\varphi'(r) c^2 \approx a_0$ within the solar
system, the value $a_0$ being imposed by the MOND regime for
$r\sim r_\text{MOND}$. But solar-system tests are precise enough
to rule out a constant anomalous acceleration even numerically as
tiny as (\ref{a0}). We must therefore look for other possible
scalar kinetic terms, and this is where generalized Galileons
enter our discussion.

Galileons were first introduced in the cosmological context in
\cite{Nicolis:2008in} (although they had actually already been
studied in \cite{Fairlie:1992} in a quite different framework).
In flat spacetime, they are theories whose field equations depend
only on second derivatives of a scalar field, but not on their
lower (0th and 1st) nor on higher derivatives. One of their
initial motivations was to generalize the key features of the
decoupling limit of the DGP brane model \cite{Dvali:2000hr},
which yields an equation of motion $\propto (\Box\varphi)^2 -
(\varphi_{;\mu\nu})^2$ for a scalar degree of freedom, playing a
crucial role in cosmology~\cite{Deffayet:2001pu,Deffayet:2000uy}.
References~\cite{Deffayet:2009wt,Deffayet:2009mn} showed how to
extend Galileon models to curved spacetime without introducing
higher derivatives (while making now first derivatives also enter
them). The same Lagrangians can be obtained in a suitable limit
of brane models including Gauss-Bonnet-Lovelock densities
\cite{deRham:2010eu}, or from dimensional reduction of such
densities \cite{VanAcoleyen:2011mj}. They can also be extended,
in any dimension, to arbitrary $p$-forms possibly coupled to each
other \cite{Deffayet:2010zh}, or to general nonlinear models
whose field equations depend on at most second derivatives
\cite{Deffayet:2011gz}. Throughout this paper, we call
Galileons this full class of models, although they go beyond
the initial ones of \cite{Nicolis:2008in}.

The new ingredient of the present paper is the third scalar
kinetic term of action (\ref{action}),
$\mathcal{L}_\text{Galileon}$, which strongly suppresses all
scalar-field effects at small distances, as we will show below.
This action (up to the factor $-8k/3$) was obtained in
Ref.~\cite{VanAcoleyen:2011mj} for the first time by dimensional
reduction of the Gauss-Bonnet-Lovelock density. It is written
here in the compact form of \cite{Deffayet:2009mn}, and is also
in the general classes considered in
\cite{Deffayet:2010zh,Deffayet:2011gz}. It is easy to check that
all field equations involve at most second derivatives, because
of the antisymmetry of the Levi-Civita tensors entering
(\ref{Galileon}).

\begin{figure}
\includegraphics[width=8.5cm]{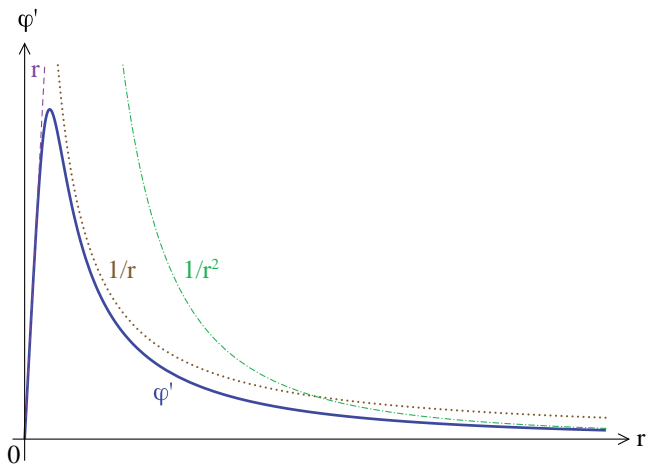}
\caption{Behavior of the scalar force $\varphi' c^2$ in the three
regimes of the theory: $\varphi' \propto r$ at small distances
(k-mouflage mechanism), $\varphi' \propto 1/r$ at intermediate
range (MOND regime), and $\varphi' \propto 1/r^2$ at large
distances (asymptotic Brans-Dicke theory).\label{fig:PhiPrime}}
\end{figure}

Although it is straightforward to write the full field equations
deriving from action (\ref{action}), for both $\varphi$ and the
metric, it will suffice in the present paper to obtain the
perturbative solution for a static and spherically symmetric
$\varphi$ in a Schwarzschild background $ds^2 = -(1-r_s/r) c^2
dt^2 + dr^2/(1-r_s/r) + r^2 d\Omega^2$, where $r_s \equiv
2GM/c^2$ denotes the Schwarzschild radius. It can be checked a
posteriori that the backreaction of the scalar field on the
metric has negligible effect as compared to present experimental
bounds. We will even assume that $r\gg r_s$ to simplify the
expressions. More detail will be provided in a forthcoming
publication \cite{BDESZ}. Denoting as before $\varphi' \equiv
d\varphi/dr$, and integrating once the field equation for
$\varphi$, we find
\begin{equation}
4 k \frac{r_s}{r^2} \varphi'^2
+ \frac{r^2 c^2}{a_0}\varphi'^2
+ \epsilon\, r^2 \varphi' \approx \frac{r_s}{2},
\label{EOM}
\end{equation}
where the origin of the different terms is obvious from
(\ref{standard})--(\ref{Galileon}). The large number of field
derivatives involved in $\mathcal{L}_\text{Galileon}$,
Eq.~(\ref{Galileon}), is responsible for the negative power of
$r$ in the first term of (\ref{EOM}). This is the central idea of
the Vainshtein mechanism, since it makes this first term dominate
at small distances. Imposing now that $\varphi'\rightarrow 0$ for
$r\rightarrow\infty$ (otherwise $\varphi$ would diverge at
infinity), we can immediately write the unique solution of
(\ref{EOM}), which is a mere second-order polynomial equation for
$\varphi'$:
\begin{equation}
\varphi' = \left(\sqrt{\frac{8 k}{r^2}+\frac{r^2 c^4}{GM a_0}
+\left(\frac{\epsilon\, r^2}{r_s}\right)^2}
+\frac{\epsilon\, r^2}{r_s}\right)^{-1}.
\label{phiprimeSol}
\end{equation}
We thus easily recover the asymptotic Brans-Dicke behavior
$\varphi' \approx GM/\epsilon\, r^2 c^2$ for very large distances
$r$, the MOND regime $\varphi' \approx \sqrt{GM a_0}/rc^2$ at
intermediate scales, but a small derivative $\varphi' \approx
r/\sqrt{8k}$ at small distances. Note that this small-distance
behavior of $\varphi'$ does not depend at all on the mass $M$,
which not only generates the right-hand side (source) of
Eq.~(\ref{EOM}), but also the background Schwarzschild geometry
entering the first term of (\ref{EOM}) via the Riemann tensor of
(\ref{Galileon}). This universal small-distance behavior means
that, paradoxically, any body generates strictly the same scalar
force $\varphi' c^2$ in its vicinity. Figure~\ref{fig:PhiPrime}
illustrates the three regimes of solution (\ref{phiprimeSol}).
Neglecting its $\epsilon$ contribution, the maximum value
$\varphi'_\text{max} = (GM a_0/2k)^{1/4}/2c$ is reached at
$r_\text{V} = (8kGM a_0)^{1/4}/c$, which defines thus a
transition radius.

In order to predict the MOND phenomenology in a galaxy of
baryonic mass $M$, we need $r_\text{V} < r_\text{MOND}$, i.e., $k
< GM c^4/8a_0^3$. The lightest dwarf galaxies for which we have
evidence for dark matter or MOND effects give thus an upper bound
for the constant $k$. Draco-like dwarfs correspond to a baryonic
mass $M$ between $10^5$ and $10^6$ solar masses, but even tiny
clusters of only $10^3 M_\odot$ seem to be dominated by dark
matter~\cite{Wolf:2009tu}. Let us thus be conservative and impose
$k \approx G(10^3 M_\odot) c^4/8a_0^3$. This is the numerical
value given in Eq.~(\ref{k}) above.

We can now estimate the order of magnitude of scalar effects
within the solar system. Since we chose a physical metric of the
disformal form (\ref{gtilde}), like in TeVeS, one could naively
conclude that the PPN parameters $\beta$ and $\gamma$ keep their
general relativistic values $\beta = \gamma = 1$. However, there
is now no meaning to define such parameters, because the PPN
formalism assumes that no length scale enters the theory, and it
needs the gravitational potential to be $\propto 1/r$ at the
Newtonian order. Therefore, the consequences of our anomalous
potential $\varphi \propto r^2 + \text{const.}$ at small
distances cannot be analyzed in the standard way. Let us thus
merely compare the anomalous force $\varphi' c^2 \approx r
c^2/\sqrt{8k}$ it generates on a test particle, with the
post-Newtonian forces $\sim (GM_\odot/r^2) \times(r_s/r)$ which
have been precisely tested in the solar system. Since their ratio
is $(r^2/r_s)^2/\sqrt{2k} \approx (r/22\,\text{AU})^4$ for the
chosen value (\ref{k}) of the constant $k$, the scalar effects
are thus smaller than $10^{-5}$ times post-Newtonian ones at the
Earth's orbit, i.e., negligible with respect to the best
experimental constraints. This ratio becomes even smaller for
inner planets, and is $< 10^{-7}$ for Mercury's orbit. On the
other hand, scalar effects grow for outer planets, but they
remain $2\times 10^{-5}$ smaller than post-Newtonian forces at
the orbit of Mars, and $3\times 10^{-3}$ at Jupiter's. This is
consistent with the most precise planetary data. When considering
the Moon's orbit, the contribution of the Earth to the Riemann
tensor entering (\ref{Galileon}) dominates over that of the Sun,
and $\varphi$ is almost spherically symmetric with respect to the
Earth's center. Denoting now as $r$ the distance to this center,
we get again the universal behavior $\varphi' c^2 \approx r
c^2/\sqrt{8k}$, which must be compared to the Newtonian
accelerations caused by both the Earth and the Sun and their
post-Newtonian corrections. We find that scalar effects on the
Moon's motion are $10^{-8}$ smaller than post-Newtonian ones,
i.e., four orders of magnitude smaller than the tightest
experimental constraints derived from Lunar Laser Ranging
\cite{Will:2005va}.

Preferred-frame effects can be estimated in a similar way.
Assuming that the solar system is moving with a velocity $\bm{w}$
with respect to the preferred frame where $U_\mu = (1,0,0,0)$, we
compute the contributions proportional to $\bm{w}^2$ in $\tilde
g_{00}$ and to $w^i$ in $\tilde g_{0i}$, and their radial
derivatives give us the magnitude of the anomalous scalar forces.
Comparing them to those generated by the $\alpha_1$-term in the
PPN formalism \cite{Will:2005va} (while the terms corresponding
to $\alpha_2$ and $\alpha_3$ vanish in the present model
\cite{Sanders:1996wk}), we find that their ratio is $\alt 8
r^3/(r_s \alpha_1\sqrt{2k}) \approx (r/10^4
\text{AU})^3/\alpha_1$, giving thus scalar effects similar to
those of an $\alpha_1 \approx 2\times 10^{-12}$ at the orbit of
Mars. As above, when considering the Moon's orbit around the
Earth, the local value of $\varphi' \approx r/\sqrt{8k}$ must be
used ($r$ denoting now the Earth-Moon distance), and we find that
preferred-frame effects caused by the scalar field are similar to
those of an $\alpha_1 \approx 10^{-15}$. Since the tightest
constraint $\alpha_1 < 10^{-4}$ comes from Lunar Laser Ranging,
we conclude that the model (\ref{action})--(\ref{gtilde}) does
not predict any detectable violation of local Lorentz invariance
in the solar system.

Binary-pulsar tests are much more subtle to compute. A precise
analysis would need either to study the (time-dependent) dynamics
of a binary system at least up to order $\mathcal{O}(1/c^3)$, or
to be able to relate scalar multipoles at infinity to their local
matter sources in spite of the nonlinearities of the Vainshtein
mechanism. In the present paper, we shall only estimate the rough
order of magnitude of scalar radiation by comparing the local
scalar forces to those of the precisely studied Brans-Dicke-like
theories. Let us first note that the monopolar radiation, naively
of order $\mathcal{O}(1/c)$, is actually reduced to order
$\mathcal{O}(1/c^5)$ because the local scalar solution generated
by any body does not depend on time; this is similar to the case
of standard scalar-tensor theories \cite{Damour:1992we}. On the
other hand, the dipolar radiation starts at order
$\mathcal{O}(1/c^3)$ in spite of the fact that the local scalar
solution is the same around any body, independently of its mass.
Indeed, if the two bodies of a binary system do not have the same
mass (say, $m_A\neq m_B)$, they do not move on the same orbit
around their common center of mass, and the global scalar field
they generate defines a preferred (oriented) direction in space.
The dominant scalar radiation is thus a dipole, and we can
estimate its order of magnitude by multiplying the one predicted
in standard scalar-tensor theories (see, e.g. Eq.~(6.52b) of
Ref.~\cite{Damour:1992we}) by the square of the ratio of the
present scalar force between the two bodies ($r c^2/\sqrt{8k}$)
and the standard one. We get that the scalar field contribution
to the time derivative of the orbital period is of order $\dot P
\sim -G c P^3 (m_A-m_B)^2(m_A+m_B)/(32 \pi^2 k m_A m_B)$, and
numerically at least $10^{-32}$ smaller than the tightest
experimental uncertainties. Although this estimate might be
erroneous by some large numerical coefficient, we can anyway
conclude that the present model should easily pass all
binary-pulsar tests.

Although the Galileon field equations involve at most second
derivatives, and avoid thus the generic instability related to
higher derivatives, this does not suffice to prove that these
models are stable. One should carefully analyze both the
boundedness by below of their Hamiltonian density and the
well-posedness of their Cauchy problem. The Hamiltonian of
flat-space Galileons is straightforward to derive
\cite{Zhou:2010di,BDESZ}, but the situation is much more complex
in curved spacetime, because all field equations for $\varphi$
and $g_{\mu\nu}$ involve second derivatives of both of them.
Around a given background, one should thus diagonalize the
kinetic terms in order to test the stability of perturbations and
the hyperbolicity of their field equations. For instance, it
would have no meaning to freeze $g_{\mu\nu}$ and study the
perturbations of $\varphi$ only in the scalar field equation
(similarly to Brans-Dicke theory, which seems to contain a ghost
scalar field for $-3/2 < \omega_\text{BD} < 0$ if one freezes the
Jordan metric, whereas studying simultaneously the dynamics of
$g_{\mu\nu}$ and $\varphi$ shows that the theory is stable even
for such a slightly negative $\omega_\text{BD}$). This problem of
the consistency of Galileon field theories goes thus beyond the
scope of this paper, and we postpone it to a forthcoming
publication \cite{BDESZ}.

Let us finally comment on our choice of Lagrangian
(\ref{Galileon}) to obtain a k-mouflage mechanism reducing scalar
effects at small distances. It happens to be the most efficient
one amongst all those that we have analyzed. The highest-order
Galileon Lagrangian which is nontrivial in 4-dimensional flat
space \cite{Nicolis:2008in,Deffayet:2009wt,Deffayet:2009mn},
namely $\mathcal{L}_5 =
-k_5\, \varepsilon^{\alpha \beta \gamma \delta}
\varepsilon^{\mu \nu \rho \sigma}
\varphi_{,\alpha} \varphi_{,\mu} \varphi_{;\beta\nu}
\left[\varphi_{;\gamma\rho} \varphi_{;\delta \sigma}
-\frac{3}{4} (\varphi_{,\lambda})^2
R_{\gamma\delta \rho \sigma}\right]$,
also generates a small $\varphi'(r) \approx \sqrt{r}/(84\,
k_5)^{1/4}$ at small distances, but its slower radial dependence
makes scalar effects still marginally detectable in the solar
system (while giving again fully negligible preferred-frame
effects). Imposing as above that the MOND phenomenology should
occur in the lightest known dwarf galaxies, we get that scalar
effects in the solar system are $\sim (r/6\,\text{AU})^{7/2}$
times post-Newtonian forces, i.e., about the order of magnitude
of the most precise bounds. Similarly, scalar effects on the
Moon's motion are $10^{-4}$ smaller than post-Newtonian ones,
i.e., of the order of current limits. A full fit of planetary
data taking into account the possible presence of such a small
scalar force would thus be necessary to test whether it is
already excluded or not. If not, this opens the exciting
possibility to detect them in future higher-precision
solar-system observations. All the other known covariant Galileon
actions are either total derivatives in 4 dimensions, or yield a
vanishing scalar field equation around a Schwarzschild background
(like Eq.~(23) of Ref.~\cite{VanAcoleyen:2011mj}), or predict a
negative $\varphi''(r)$ and therefore too large scalar forces in
the solar system. On the other hand, generalized Galileon actions
\cite{Deffayet:2011gz} involving non-differentiated fields
$\varphi$ and/or negative powers of $s\equiv
(\varphi_{,\lambda})^2$ can provide alternative models
\cite{BDESZ}, but anyway less natural than (\ref{Galileon}).



\end{document}